\documentclass[aps,prl,twocolumn,epsfig,amsmath,amssymb,showpacs]{revtex4}
\usepackage[english]{babel} \usepackage{latexsym}
\usepackage{graphicx} \usepackage{subfigure} \usepackage{epsfig}
\usepackage{psfrag}

\usepackage{amssymb}
\usepackage{amsmath}
\usepackage{amscd}
\usepackage{here}
\usepackage{bbm}

\begin{document}

\title{Hybrid multi-site excitations in dipolar condensates in optical lattices}
\author{M. Klawunn and L. Santos}
\affiliation{
\mbox{Institut f\"ur Theoretische Physik , Leibniz Universit\"at
Hannover, Appelstr. 2, D-30167, Hannover, Germany}}

\begin{abstract}
Strong 1D lattices usually lead to unconnected 
two-dimensional gases. The long-range character of the dipole-dipole 
interactions leads to a novel scenario where non-overlapping 
gases at different sites may interact significantly.  
We show that the excitations of non-overlapping condensates in 1D optical 
lattices acquire a band-like character, 
being collectively shared by different sites. In particular, the 
hybridization of the modes significantly enhances the 
rotonization of the excitations, and may induce roton-instability. 
We discuss the observability of this effect in on-going experiments.
\end{abstract}

 \maketitle



A novel path on cold gases is currently being opened by recent experiments in which  
(magnetic or electric) dipole-dipole interaction (DDI) 
plays a significant or even dominant role. 
On one side the recent creation of heteronuclear 
molecules in the lowest ro-vibrational level 
\cite{Ospelkaus2008,Deiglmayr2008} 
opens exciting perspectives for the achievement of 
a quantum degenerate gas of polar molecules, which may possess 
large dipole moments (e.g. $\sim 0.5$ Debyes for KRb \cite{Ospelkaus2008}). 
On the other side, the magnetic DDI has been shown to lead to 
exciting novel phenomena in recent experiments in Bose-Einstein 
condensates (BECs) of Chromium (which has a magnetic moment 
$\mu=6\mu_B$, with $\mu_B$ the Bohr magneton) \cite{Chromium}. Particularly interesting 
is the fact that the short-range interactions (SRI) may be suppressed by means of Feshbach resonances, 
leading to a purely dipolar gas \cite{Feshbach-Tilman}. The 
DDI plays also a significant role in very recent experiments 
on spinor Rubidium BECs, in spite of the small $\mu=1\mu_B$, since the energy 
scale of the DDI becomes comparable with the (also very small) energy 
scale of spin-changing collisions \cite{Vengalattore2008}. 
Very recent experiments
have shown as well that the DDI leads to a observable   
damping of Bloch oscillations in Potassium BECs in tilted optical lattices 
\cite{Fattori2008}.


The partially attractive character of the DDI results in nontrivial
stability conditions for dipolar BECs. Low-momentum instability (phonon instability) 
\cite{Santos2000} results in a geometry-dependent 
instability against collapse in 3D traps, as recently observed experimentally 
\cite{CollapsePfau}, or soliton formation in 2D geometries \cite{Nath2008}. 
Interestingly, the momentum dependence of the DDI  
allows for a second type of instability (roton instability) related 
to the appearance of a roton-like minimum in the dispersion law 
of elementary excitations \cite{Roton}. Roton instability leads to local 
collapses \cite{Komineas} or stabilized modulated density profiles in
sufficiently tight traps \cite{Ronen}.


The long-range character of the DDI induces a nonlocal nonlinearity in dipolar BECs
that resembles that encountered in plasmas \cite{Plasmas} or 
nematic liquid crystals \cite{nematics}. As a consequence, novel 
phenomena as stable 2D solitons become possible 
\cite{Pedri2005,Vardi2007}. This nonlocality leads to fundamentally new
physics for quantum gases in optical lattices, since it induces interactions 
between neighboring sites. As a consequence, dipolar bosons in optical
lattices are described by extended versions of the Bose-Hubbard Hamiltonian, and may 
present a wealth of novel phases, as supersolid \cite{Goral2002} or
Haldane-phases \cite{DallaTorre2006}. In addition, contrary to the case of
SRI, very deep optical lattices do not lead to independent low-dimensional
gases, since non-overlapping atoms at different sites interact. 
As a consequence nonoverlapping BECs in two-well potentials may 
scatter \cite{Nath2007}, pair superfluidity may appear in ladder-like lattices 
\cite{Arguelles2007}, and even filament condensation may occur
\cite{Wang2006}. The effects of the intersite DDI have been observed 
experimentally for the first time in very recent experiments in Florence on 
Bloch oscillations \cite{Fattori2008}.

\begin{figure}
\begin{center}
\includegraphics[width=4.8cm]{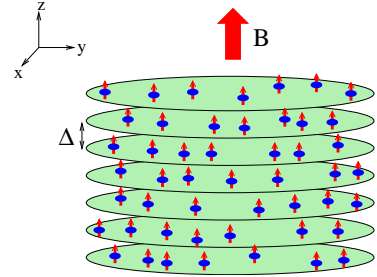}
\end{center}
\vspace*{-0.5cm}
\caption{Scheme of the system under consideration.}
\vspace*{-0.2cm}
\label{fig:1}
\end{figure}


This Letter is devoted to the analysis of non-overlapping dipolar BECs placed 
at different sites of a deep two-well potential or 1D optical lattice. 
As mentioned above, contrary to the case of purely SRI, the deep potential 
does not lead to independent 2D BECs.  In particular, we show that
the elementary Bogoliubov excitations of disconnected BECs placed in 
a two-well potential couple through the DDI leading 
to hybrid modes which are collectively shared by both wells. Interestingly 
this hybridization may significantly alter the stability of the system against 
roton instability. We show that this effect is significantly 
enhanced for the case of a 1D optical lattice with multiple sites, where 
a band-like spectrum is induced by the inter-site DDI. 
We analyze in detail the stability
diagram, and finish with a discussion of the experimental observability in
different experiments.


In the following, we consider a BEC of particles with mass $m$ and 
electric dipole $d$ (the results are equally valid for magnetic dipoles) 
oriented along $z$ by an external field, and 
that hence interact via a dipole-dipole potential: 
$V_d(\vec{r})= d^2 (1-3\cos^2(\theta))/r^3$, where $\theta$ 
is the angle formed by $\vec r$ with the $z$ axis. 
At sufficiently low temperatures the physics 
of the dipolar BEC is provided by a non-local non-linear Schr\"odinger 
equation (NLSE):
\begin{eqnarray}
&&i\hbar\frac{\partial}{\partial t}\Psi(\vec r,t)=\left [
-\frac{\hbar^2}{2m}\nabla^2+V(\vec r)
+g|\Psi(\vec r,t)|^2 \right\delimiter 0 \nonumber \\
&&+ \left\delimiter 0  \int d^3 r' V_d(\vec r-\vec r')
|\Psi(\vec r',t)|^2
\right ]\Psi(\vec r,t),
\label{GPE}
\end{eqnarray}
where $g=4\pi\hbar^2a/m$, with $a$ the $s$-wave scattering length,  
and $V(\vec r)$ is the trap potential. 
In the following we use the convenient dimensionless parameter $\beta=g_d/g$, 
that characterizes the strength of DDI compared to the short 
range interaction, where $g_d=8\pi d^2/3$. Note that $\beta$ 
may be easily controlled experimentally by means of Feshbach 
resonances, as recently shown in Ref.~\cite{CollapsePfau}.

\begin{figure}[ht]
\begin{center}
\includegraphics[width=4.8cm]{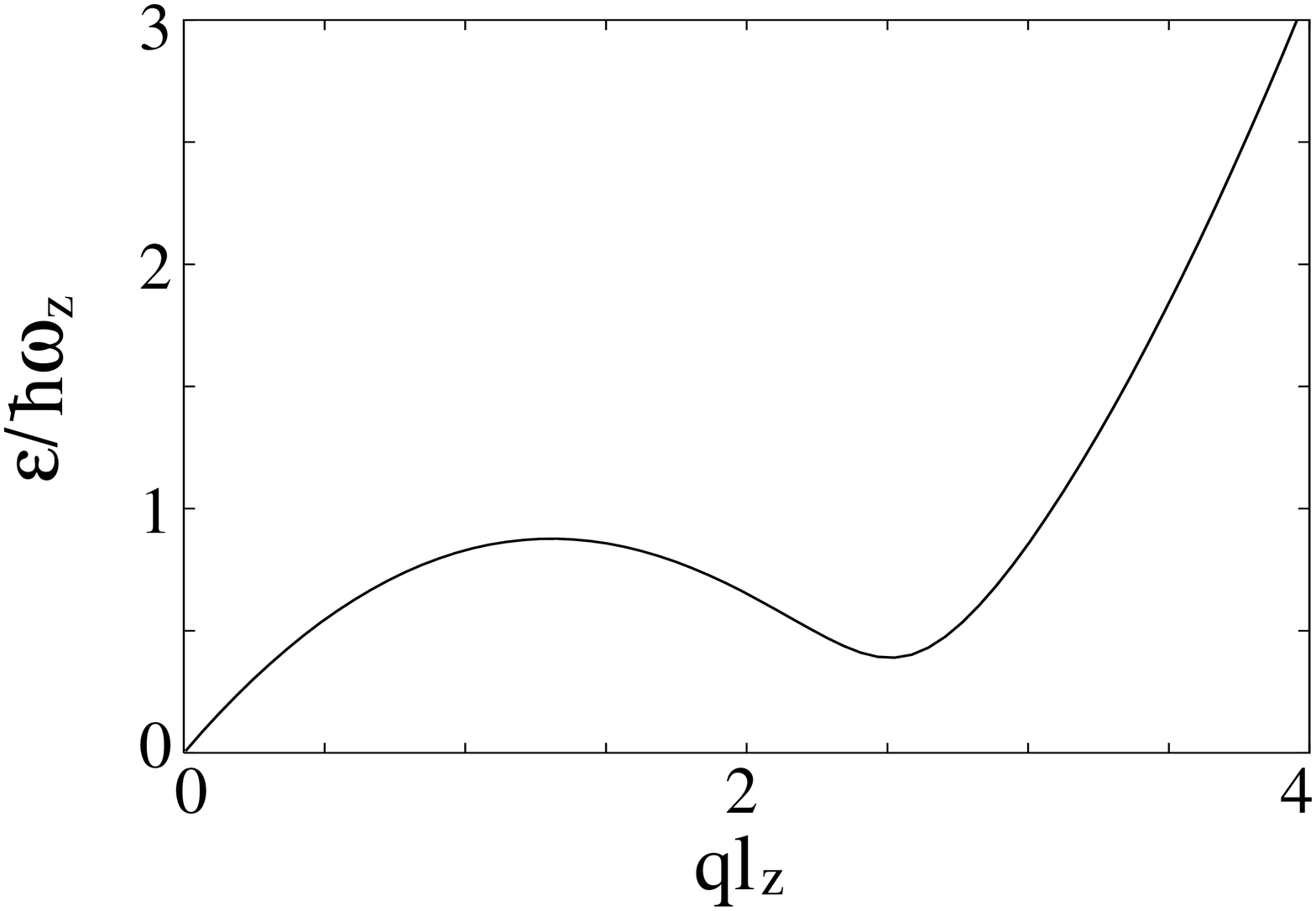}
\end{center}
\vspace*{-0.5cm}
\caption{Roton minimum in the dispersion relation $\epsilon/\hbar\omega_z$
of a single 2D BEC with $\beta=-1.07$, $l_z=0.09\mu$m, $a=-2$ nm, and 
a 3D density of $n_0/\sqrt{2\pi}l_z=10^{14}{\rm cm}^{-3}$.}
\vspace*{-0.2cm}
\label{fig:2}
\end{figure}

We consider first the case of a quasi-2D homogeneous BEC confined in $z$ by  
$V(\vec r)=m\omega_z^2 z^2/2$, which is 
sufficiently strong such that $\Psi(\vec r,t)=\Psi_\perp(\vec\rho,t)\Phi_0(z)$,
where $\Phi_0(z)=\exp(-z^2/2l_z^2)/\pi^{1/4}l_z^{1/2}$ is the ground
state of the transversal oscillator ($l_z=\sqrt{\hbar/m\omega_z}$). 
The ground state of the homogeneous 2D BEC is of the form 
$\Psi_\perp(\vec\rho,t)=\exp(-i(\mu/\hbar+\omega_z)t)\sqrt{n_0}$, 
where $n_0$ is the 2D density, and
$\mu$ is the 2D chemical potential.  Introducing 
this form into the NLSE~(\ref{GPE}), one obtains
$\mu=(g+g_d)n_0/\sqrt{2\pi}l_z$ (the 2D condition is satisfied for $\mu\ll\hbar\omega_z$). 
Inserting a plane-wave Ansatz $\Psi(\vec r,t)=(\sqrt{n_0}+
u_q \exp(i\vec q\cdot\vec\rho-i\epsilon t/\hbar)-v_q^*\exp(-i\vec
q\cdot\vec\rho-i\epsilon t/\hbar))
\Phi_0(z)\exp(-i(\mu/\hbar+\omega_z)t)$ 
into the NLSE~(\ref{GPE}), and linearizing in $u_q$, $v_q$,
we obtain the Bogoliubov spectrum of elementary excitations:
\begin{equation}
\epsilon(q)=\left\{ E_q \left [ E_q + 2A \right ]\right\} ^{1/2}
\label{BG1}
\end{equation}
where $E_q=\hbar^2q^2/2m$, 
$A=\mu-(g_dn_0/\sqrt{2\pi}l_z)F(ql_z/\sqrt{2})$
and $F(x)=\frac{3\sqrt{\pi}}{2} |x|{\rm erfc}(x)e^{x^2}$. Note that 
without DDI ($\beta=0$) we recover the usual Bogoliubov spectrum for a 2D BEC
with purely SRI. In particular, if $a<0$ and $\beta=0$, $\epsilon (q)^2<0$ for 
$q\rightarrow 0$, recovering the well known phonon instability (and subsequent
collapse) in homogeneous BEC with $a<0$. If the dipole is sufficiently large,
such that $g+g_d>0$, then the DDI prevents the instability at $q\rightarrow 0$. 
However, due to the $q$-dependence of the DDI (given by the monotonously
decreasing character of the function $F$), the dispersion $\epsilon(q)$ may show for
intermediate $g_d$ values a roton-like minimum at a finite value of $q l_z$ 
(Fig.~\ref{fig:2}). For sufficiently low DDI $\epsilon(q)^2<0$ 
at the roton-like minimum, leading to dynamical instability (roton
instability). 
For $|\beta|>\beta_{cr}$ (with $\beta_{cr}$ dependent on the 
ratio $gn_0/l_z\hbar\omega_z$) roton instability is prevented, 
and the 2D homogeneous BEC is stable. 

In the following we show that
$\beta_{cr}$ is significantly modified in the presence of other neighboring 
quasi-2D dipolar BECs. We consider the case of an optical lattice 
along $z$ (Fig.~\ref{fig:1}) described by a potential  
$V(\vec r)=s E_r \sin^2(\pi z/\Delta)$, where $\Delta$ is the intersite 
spacing, and $s$ provides the lattice depth in units of the recoil energy 
$E_r=\hbar^2\pi^2/2m\Delta^2$. As in the previous discussion we consider 
no trapping on the $xy$-plane (we discuss the potentially important role 
of the harmonic confinement on the $xy$-plane at the end of this Letter).
At each lattice node $V(\vec r)$ may be approximated by 
an effective harmonic oscillator potential $V_{eff}(z)$, 
with effective oscillator length $l_z\approx \Delta s^{-1/4}/ \pi$. 
The lattice is considered strong enough so that we can assume that there is
no spatial overlap between wave-functions in different lattice sites, and 
hence we may neglect any hopping.
We assume the DDI small enough to neglect pairing \cite{Arguelles2007} or filamentation
\cite{Wang2006}.


We start our discussion on inter-site effects with the two-well case. 
This simplified scenario already captures many features of
the effect discussed. In addition, two-well potentials may be experimentally 
realized and are currently of considerable interest 
\cite{Schmiedmayer,Oberthaler}. The quasi-2D BEC in the $i$-th layer
is given by the extended NLSE:

\begin{eqnarray}
&&i\hbar\frac{\partial}{\partial t}\Psi_i(\vec r,t)=\left [
-\frac{\hbar^2}{2m}\nabla^2+V_{eff}(z)
+g|\Psi_i(\vec r,t)|^2 \right\delimiter 0 \nonumber \\
&&+ \left\delimiter 0  \sum_j \int d^3 r' V_d(\vec r-\vec r')
|\Psi_j(\vec r',t)|^2
\right ]\Psi(\vec r,t),
\label{GPE2}
\end{eqnarray}
where for a two-well potential, $i,j =1,2$. Note that, crucially,  
the DDI couples now the $i$-th layer to the $j$-th one.
Similar to the single-site discussion, we consider a strong $z$-confinement 
at each site, and hence we may employ a quasi-2D Ansatz 
$\Psi_i(\vec{r})=\Psi_{\perp,i}(\vec{\rho},t)\Phi_0(z-z_i)$, 
where $\Phi_0(z)$ has the form discussed above, and $z_i$ 
is the position of the $i$-th lattice node. The ground-state 
of the condensates at the two layers in given by the Ansatz  
$\Psi_{\perp,i}(\vec{\rho},t)=\sqrt{n_0} e^{-i(\mu/\hbar+\omega_z)t}$, 
where we consider the same 2D density $n_0$ at both sites. 
Introducing this Ansatz into the NLSE~(\ref{GPE2}) we obtain 
the 2D chemical potential 
$\tilde\mu=\mu+\lambda(\Delta)$, with $\mu$ the chemical potential 
of an individual well and 
$\lambda(\Delta)=(g_d n_0/\sqrt{2\pi} l_z ) e^{-\Delta^2/l_z^2} $

As above we are interested in the elementary excitation of these systems. 
For $\Delta\rightarrow\infty$ the Bogoliubov modes at each site 
are independent and described by the single-site expression~(\ref{BG1}). 
For finite $\Delta$ the inter-site coupling leads to an hybridization of the 
modes at both sites with significant consequences, as we discuss below. 
As for the single-site discussion we insert 
a plane-wave Ansatz $\Psi_i(\vec r,t)=(\sqrt{n_0}+
u_{qi} \exp(i\vec q\cdot\vec\rho-i\epsilon t/\hbar)-v_{qi}^*\exp(-i\vec
q\cdot\vec\rho-i\epsilon t/\hbar))
\Phi_0(z)\exp(-i(\tilde\mu/\hbar+\omega_z)t)$ 
into the NLSE~(\ref{GPE2}), and linearize in $u_{qi}$, $v_{qi}$. 
In this way we obtain four coupled Bogoliubov-de Gennes equations 
for $\{ u_{1,2},v_{1,2} \}$, which may be diagonalized to 
obtain the Bogoliubov modes:
\begin{equation}
\epsilon_{\pm}(q)=\left\{ E_q \left [E_q+2A\pm C(\Delta) \right ] \right\}^{1/2},
\label{BG2}
\end{equation}
where 
\begin{equation}
C(\Delta)=\lambda(\Delta)-\frac{g_d n_0}{\sqrt{2\pi}l_z}
\tilde F\left (\frac{ql_z}{\sqrt{2}},\frac{\Delta}{\sqrt{2}l_z} \right ),
\label{C}
\end{equation}
with 
$
\tilde F(x,y)=\frac{3\sqrt{\pi}xe^{x^2}}{4}\sum_{\alpha=\pm 1} 
e^{-2\alpha xy}{\rm erfc}(x-\alpha y).
$

\begin{figure}[ht]
\begin{center}
\includegraphics[width=7cm]{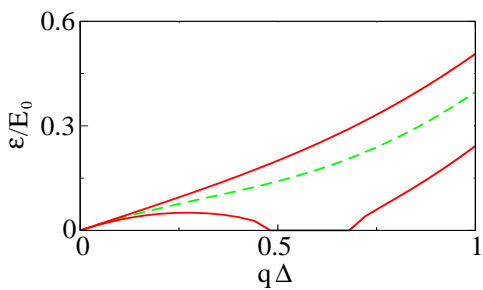}
\end{center}
\vspace*{-0.5cm}
\caption{Dispersion law (in units of $E_0=\hbar/m\Delta^2$) 
for a single site (dashed) and 
for two wells (solid) for $\beta=-1.2$, 
$\Delta=0.53$ $\mu$m, $s=13.3$,
$a=-2$ nm, and $n_0/\sqrt{2\pi}l_z=10^{14}/{\textrm{cm}}^3$.
}
\vspace*{-0.2cm}
\label{fig:3}
\end{figure}

Note that for $\Delta\rightarrow\infty$, $C(\Delta)=0$ and  
we recover two degenerate independent modes. 
For finite $\Delta$ the modes at the two wells 
hybridize, and two different branches appear
for each $q$, one stiffer than the modes
for $\Delta\rightarrow\infty$, and the other softer. The latter is
particularly interesting, since the soft mode is 
more prone to rotonization (Fig.~\ref{fig:3}). 
Interestingly, under proper conditions, two parallel 
non-overlapping BECs may become roton-unstable even if 
they were stable separately. 
As a consequence, a larger $\beta_{cr}$ is necessary to stabilize 
the two-well system.

\begin{figure}
\begin{center}
\includegraphics[width=5.5cm]{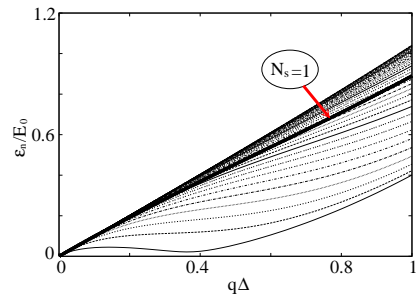}
\end{center}
\vspace*{-0.5cm}
\caption{Band-like dispersion (in units of $E_0=\hbar/m\Delta^2$)
for $N_s=40$ and $\beta=-2.44$.
Other parameters are as in Fig.~\ref{fig:3}. We indicate the dispersion 
law for $N_s=1$.}
\vspace*{-0.2cm}
\label{fig:4}
\end{figure}

The hybridization (and consequent destabilization) in two-well potentials 
becomes even more pronounced for the case of dipolar BECs at $N_s>2$ 
sites of a 1D optical lattice, since a site $i$ couples with 
all its neighbors $j$ 
(of course with decreasing strength for growing $|i-j|$). For simplicity 
of our analysis we consider the case in which all lattice sites present 
the same 2D density $n_0$.   
In that case, one may easily generalize the two-site analysis to 
the multi-site case, to reach a set of coupled Bogoliubov-de Gennes 
equation for $f_{qi}=u_{qi}+v_{qi}$:
\begin{equation}
\epsilon^2 f_{qi}=E_q(E_q+2A)f_{qi}+2E_q \sum_{j\neq i} C(\Delta |i-j|) f_{qj}.
\label{BdG}
\end{equation}
After diagonalizing the matrix of coefficients at the rhs of Eqs.~(\ref{BdG}), 
we obtain numerically the corresponding band-like set of 
$N_s$ elementary excitations (Fig.~\ref{fig:4}). Note that 
the band-like spectrum has an upper
phonon-like boundary which for large $N_s$ has an approximate 
sound velocity $c_s\simeq\sqrt{(A+\sum_n C(\Delta|n|))/m}$.  
The lower mode of the $N_s$ manifold becomes significantly softer than the 
individual modes for independent sites. As a consequence the roton instability 
extends to larger $\beta_{cr}$ when $N_s$ increases, 
until saturating for a sufficiently large $N_s$ (due to the 
decreasing DDI for increasing distance between sites). 

\begin{figure}[ht]
\begin{center}
\includegraphics[width=6.0cm]{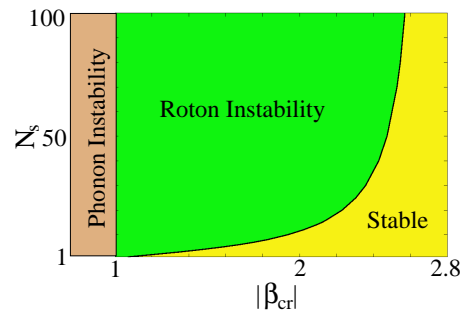}
\end{center}
\vspace*{-0.5cm}
\caption{
Stable and unstable regimes for $N_s$ dipolar 2D BECs.
We employ the same parameters as in  Fig.~\ref{fig:3}.}
\vspace*{-0.2cm}
\label{fig:5}
\end{figure}


Fig.~\ref{fig:5} summarizes our results on the stability 
as a function of $\beta$ (we recall that $g<0$). As mentioned above if 
$g+g_d<0$ ($|\beta|<1$) the system is unstable against phonon instability. For 
$1<|\beta|<|\beta_{cr}(N_s)|$ the system is unstable against 
roton instability. $|\beta_{cr}|$ increases when $N_s$ grows 
until saturating for
sufficiently large $N_s$. For $|\beta|>|\beta_{cr}(N_s)|$ the  
quasi-2D BECs are stable.  

\begin{figure}[ht]
\begin{center}
\includegraphics[width=5.5cm]{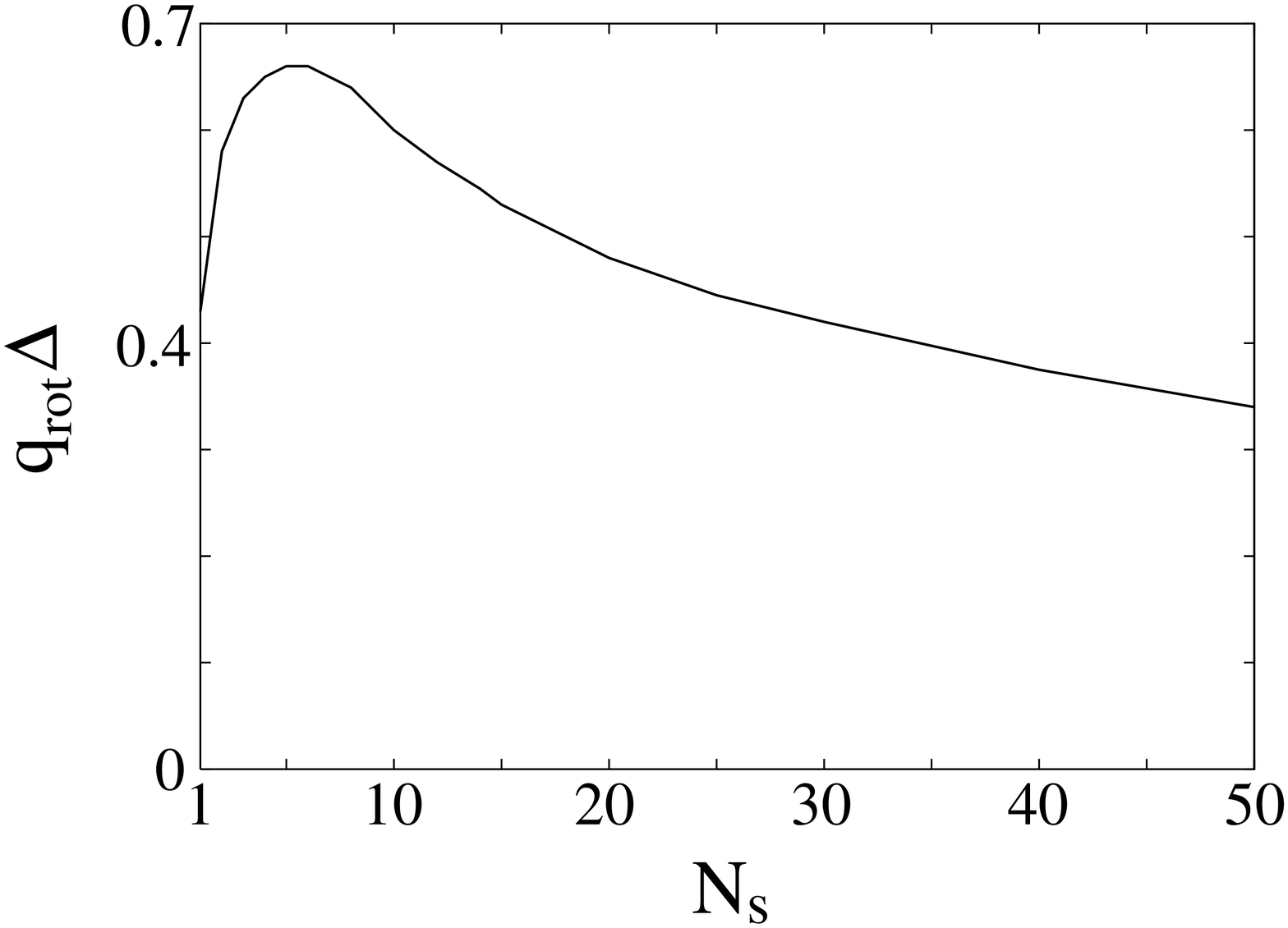}
\end{center}
\vspace*{-0.5cm}
\caption{Roton momentum $q_{rot}$ when the roton minimum touches zero 
as a function of the number of sites $N_s$.
We employ the same parameters as in  Fig.~\ref{fig:3}.
}
\vspace*{-0.2cm}
\label{fig:6}
\end{figure}

 
The value of $q_{rot}$ when the roton becomes unstable 
is of particular importance. Fig.~\ref{fig:6} shows a typical variation of 
$q_{rot}\Delta$ at the curve $\beta=\beta_{cr}(N_s)$ as a function of $N_s$.
Note that $q_{rot}$ at $\beta_{cr}$ shows a maximum for small $N_s$. 
For small $N_s$, $\beta_{cr}$ (and hence the on-site repulsive DDI) 
increases significantly, and hence the value of $q_{rot}$ at 
$\beta_{cr}$ increases. For larger $N_s$ $\beta_{cr}$
tends to saturate, as mentioned above, and the repulsive on-site DDI 
remains approximately constant along the curve $\beta_{cr}(N_s)$. 
As a consequence, the increase in $N_s$ just increases 
the attractive contribution of the DDI of neighboring sites, and the 
DDI becomes less effective in compensating the attractive on-site SRI. 
As a result of that, $q_{rot}$ decreases until saturating at a value lower 
than that for a single site.  

Typical experiments work with an harmonic $xy$-trapping 
(of frequency $\omega_{xy}$). 
Although we have assumed homogeneous quasi-2D gases, 
we may estimate the effect of the $xy$-trapping by 
considering an effective cut-off at low momenta 
$q_{cut}\simeq 1/l_{xy}$, where 
$l_{xy}=\sqrt{\hbar/m\omega_{xy}}$ is the harmonic oscillator length 
characterizing the $xy$-trap. In a good approximation we may consider 
that all features occurring at momenta $q<q_{cut}$ are suppressed 
by the trap. As a consequence one expects that the $xy$ confinement 
suppresses roton instability for frequencies $\omega_{xy}>\omega_{cut}$.
For typical densities $10^{14} {\rm cm}^{-3}$, and typical intersite 
separation $\Delta=0.53\mu$m, we estimate for $^{52}$Cr 
that for a single site $\beta_{cr}$ is achieved at a 
scattering length $a\simeq -31 a_0$, and that for this case 
$\omega_{cut}\simeq 66$Hz. For the same case but 
$N_s=4$ (which is the maximum of the corresponding $q_{rot}$ curve), 
$\beta_{cr}$ is achieved for 
$a\simeq -24 a_0$, and $\omega_{cut}\simeq 160$Hz. 
For the latter case an instability rate of 
$\Gamma^{-1}\simeq 5$ms is expected for $a=-24.5 a_0$.  
For $^{39}$K, the numbers are more restrictive 
(due to the lower magnetic moment). For $N_s=25$ (maximum 
of the $q_{rot}$ curve), $\beta_{cr}$ is achieved 
for $a\simeq -0.52 a_0$, and $\omega_{cut}\simeq 4$Hz. 
For this case, one expects $\Gamma^{-1}\simeq 180ms$ 
at $a=-0.53a_0$.

Note that the fact that $q_{rot}$ shows a maximum may have 
interesting consequences in experiments, since this suggest that 
for some intermediate trapping frequencies the instability 
may be just present for a given window of values of $N_s$. 
Note also that the previous discussion just refers to the 
destabilization when the roton touches zero. For even larger 
values of $|a|$, a larger region of $q$ may become unstable. However 
if the $xy$-trap just allows for the resolution of the upper 
boundary of the unstable region and not the lower one, 
roton and phonon instability may become experimentally undistinguishable. 


Summarizing, the nonlocal 
character of the DDI leads 
to a novel scenario where non-overlapping 
gases at different sites interact significantly. 
Contrary to the case of pure SRI, the DDI 
leads to the hybridization of the excitations 
at different sites, which acquire a collective band-like character. 
In particular, the hybridization of the modes leads to a 
significant enhancement of the rotonization of the excitations, 
and may induce roton-instability for values of the SRI at which 
a single site is stable. Finally, we have discussed the experimental 
requirements for the observation of the roton instability.

{\em Note: } After the completion of this work we became aware of a similar 
analysis by Wang and Demler \cite{WangDemler}, 
in which roton-softening due to intersite interactions is discussed 
in the context of recent experiments in Florence \cite{Fattori2008}. 
Although we consider that roton softening plays no significant role 
in the damping observed in Ref.~\cite{Fattori2008} 
due to the $xy$-confinement, a weaker $\omega_{xy}$ 
(along the lines discussed above) could allow for the instability  
discussed by Wang and Demler, and by us in this manuscript. 
However, a more careful quantitative analysis is necessary, taking into 
account both the $xy$-trapping, and the $z$-trapping, which we 
plan to investigate in a further work.


\acknowledgements

We thank R. Nath, T. Pfau, M. Fattori, G. Modugno and M. Jona-Lasinio 
for useful discussions. This work was supported by the DFG 
(SFB407, QUEST), and the ESF (EUROQUASAR). 


\end{document}